\documentclass[a4paper]{jpconf}
\usepackage{graphicx}
\begin{document}
\title{Isolated Leptons and Missing $P_{T}$ at HERA}

\author{David M. South}

\address{Universit\"{a}t Dortmund, Experimentelle Physik V,
	44221 Dortmund, Germany}

\ead{david.south@desy.de}

\begin{abstract}
The search for events containing isolated leptons (electrons or muons)
and missing transverse momentum produced in $e^\pm p$ collisions is
performed individually and in a common phase space with the H1 and ZEUS
detectors at HERA in the period 1994--2007.
The analysed H1+ZEUS data sample corresponds to an integrated luminosity of
0.97~fb$^{-1}$, and composes of the complete high energy data from the
HERA programme.
A total of 87 events are observed in the data, compared to a
Standard Model (SM) prediction of 92.7~$\pm$~11.2.
At large hadronic transverse momentum $P_{T}^{X} >$~25~GeV in
the $e^{+}p$ data, luminosity 0.58~fb$^{-1}$, 23 data events are
observed compared to a SM prediction of 14.6~$\pm$~1.9.
Production cross section measurements of events containing isolated
leptons and missing transverse momentum and single $W$ production,
as well as $W$ polarisation fractions performed by the H1 Collaboration
are also presented.
The H1 isolated lepton events are also examined in the context of a
search for single top production.

\end{abstract}

\section{Introduction}

Events containing a high $P_{T}$ isolated electron or muon and associated
with missing transverse momentum have been observed at
HERA \cite{isoleph1origwpaper,isoleph1newwpaper,isolepzeusorigwpaper,zeustop}.
An excess of HERA~I data events (1994--2000, mostly in $e^{+}p$ collisions)
compared to the SM prediction at large hadronic transverse
momentum $P_{T}^{X}$ was reported by the H1 Collaboration
\cite{isoleph1newwpaper}, which was not confirmed by the ZEUS Collaboration,
although the analysis was performed in a more restricted phase
space \cite{zeustop}.


The main SM contribution to the signal topology is the production
of real $W$ bosons via photoproduction with subsequent leptonic decay
$ep\rightarrow eW^{\pm}$($\rightarrow l\nu$)$X$, where the hadronic
system $X$ is typically of low $P_{T}$. 
The equivalent charged current process
$ep \rightarrow \nu$$W^{\pm}$($\rightarrow l\nu$)$X$ also contributes to the
total signal rate, although only at a level of around 7\%.
The production of $Z^{0}$ bosons with subsequent decay to neutrinos
$ep \rightarrow eZ^{0}$($\rightarrow \nu\bar{\nu}$)$X$ results in a
further minor contribution\footnote{This process is not included in
the present ZEUS analysis.} to the total signal rate in the electron
channel at a level of 3\%.
%
%


The event selection employed by the H1 \cite{h1isolepnew} and ZEUS
\cite{zeusisolepnew} analyses are very similar and may be summarised
as follows:
The identified lepton should have high transverse momentum $P_{T}^{l} >$~10~GeV,
be observed in the central region of the detector and be isolated with
respect to jets and other tracks in the event.
The event should also contain a large transverse momentum imbalance,
$P_{T}^{miss} >$~12~GeV. Further cuts are then applied, which are designed
to reduce SM background, whilst preserving a high level of signal purity.
Event quantities sensitive to the presence of high energy undetected
particles in the event are employed such as the azimuthal balance of the
event, the difference in azimuthal angle between the lepton and the
hadronic system and the longitudinal momentum imbalance.
To ensure that the two lepton channels are exclusive and may therefore
be combined, electron events must contain no isolated muons.


\section{Results from the H1 and ZEUS Analyses}
\label{sec:sep}

Both H1 and ZEUS have recently performed the analysis of the electron and
muon channels\footnote{The H1 Collaboration have also performed the
analysis of the tau decay channel using the full HERA~I+II data and the
hadronic 1--prong tau decay mode \cite{h1isotaunew}. In this search, where the
signal purity is much lower at around 14\%, 20 events are observed in the
data compared to a SM prediction of 19.5~$\pm$~3.2.} on their respective
complete HERA I+II data sets, which correspond to
approximately 0.5~fb$^{-1}$ per experiment \cite{h1isolepnew,zeusisolepnew}.
A total of 59 events are observed in the H1 data, compared to a
SM prediction of 58.9~$\pm$~8.2.
For $P_{T}^{X} >$ 25~GeV, a total of 24 events are observed compared
to a SM prediction of 15.8~$\pm$~2.5, where 21 events are observed in
the $e^{+}p$ data compared to a SM prediction of 8.9~$\pm$~1.5.
The observed data excess in the HERA~I $e^{+}p$ data thus remains at
the 3.0$\sigma$ level for the complete H1 $e^{+}p$ dataset.
In the ZEUS analysis of the complete HERA~I+II data, 41 data events are
observed in agreement with the SM prediction of 48.3~$\pm$~6.8.
Unlike in the H1 analysis, agreement between data and SM is also observed
in the high $P_{T}^{X}$ region, where 11 events are seen in
the $e^{\pm}p$ data compared to a SM prediction of 13.1~$\pm$~1.8.


\section{A Combined H1 and ZEUS Analysis}
\label{sec:comb}

%

A study of the selection efficiency for signal process using the event
generator EPVEC \cite{epvec} found the H1 and ZEUS analyses
to be compatible in the kinematic region where they are directly
comparable \cite{h1ichep06,zeusichep06}.
The majority of the data events observed by H1 at $P_{T}^{X} >$ 25~GeV
are also found to fall into the region of overlap of the two analyses.
Nevertheless, in order to coherently combine the results from the two
experiments, a common phase space has been established.
The common selection is based
on the H1 event selection \cite{isoleph1newwpaper,h1isolepnew}, but over a
more restricted lepton polar angle range of
15~$^\circ < \theta_l <$~120~$^\circ$, as employed in the ZEUS analysis
\cite{zeusisolepnew}.
The signal expectation rates of the H1 and ZEUS analyses using the common
selection are found to be comparable, taking into account the
respective luminosities of the data sets and the signal processes included.
More details on the combination of the H1 and ZEUS analyses can be found
in \cite{h1andzeusisolepnew}.

\begin{figure}
  \includegraphics[height=.42\textwidth]{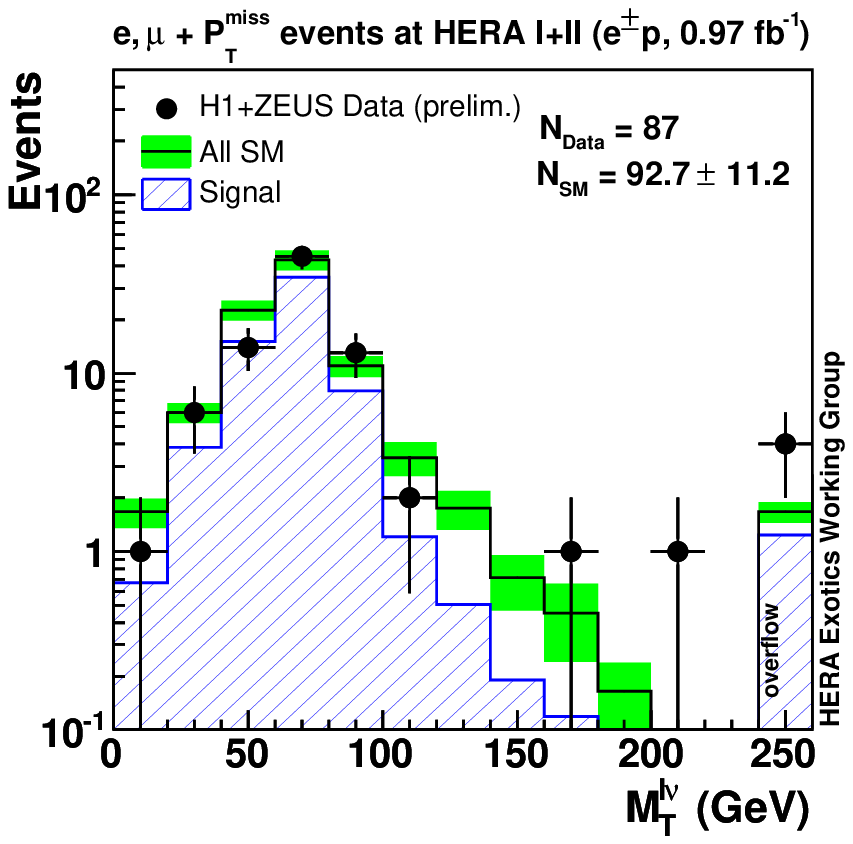}
  \hfill
  \includegraphics[height=.42\textwidth]{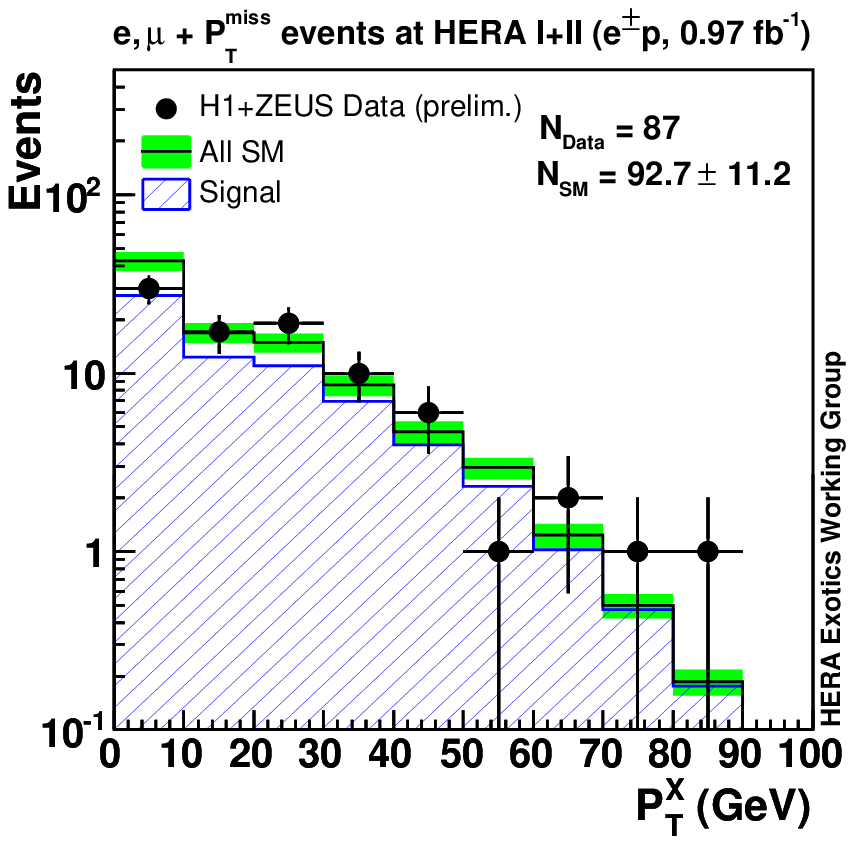}
  \caption{The transverse mass $M_{T}^{l\nu}$ (left) and hadronic
	transverse momentum $P_{T}^{X}$ (right)
	distributions of the combined H1+ZEUS $e^{\pm}p$ HERA I+II data.
	The data (points) are compared to the
	SM expectation (open histogram). The signal component of the SM
	expectation is given by the
	hatched histogram. $\rm N_{Data}$ is the total number of data events
	observed, $\rm N_{SM}$ is the total SM expectation. The total error
	on the SM expectation is given by the shaded band.}
\label{fig:isolep1}
\end{figure}

The results of the combined H1+ZEUS analysis are summarised in table 1.
The signal contribution, dominated by real $W$ production, is seen to
dominate the total SM expectation in all data samples.
At large hadronic transverse momentum $P_{T}^{X} >$ 25~GeV a total of
29 events are observed in the H1+ZEUS $e^{\pm}p$ data compared to a
SM prediction of 25.3~$\pm$~3.2.
In the $e^{+}p$ data alone, 23 events are observed with $P_{T}^{X} >$ 25~GeV
compared to a SM prediction of 14.6~$\pm$~1.9, equivalent to an excess
of data over the SM prediction of 1.8$\sigma$.
Seventeen of the 23 data events are observed in the H1 data compared to
a SM expectation of 7.1~$\pm$~0.9, equivalent to an excess of data over
the SM prediction of 2.9$\sigma$.
Figure~\ref{fig:isolep1} shows the transverse mass, $M_{T}^{l\nu}$
and $P_{T}^{X}$ distributions of the H1+ZEUS $e^{\pm}p$ data for the
combined electron and muon channels.
%
%

\section{Cross Sections and $W$ Polarisation Fractions}

The H1 selection results described in section \ref{sec:sep} are used to
calculate production cross sections for events with an energetic isolated
lepton and missing transverse momentum ($\sigma_{\ensuremath{\ell+{P}_{T}^{miss}}}$)
and for single $W$ boson production ($\sigma_{W})$, for which the branching
ratio for leptonic $W$ decay is taken into account \cite{h1wpol}.
The results are shown below with statistical (stat) and systematic (sys)
uncertainties compared to the SM, quoted with a theoretical systematic error
(th.sys) of 15\%.

\vspace{-0.2cm}

\begin{table}[h]
\begin{center}
\begin{tabular}{ | l | c @{$\,\pm\,$} c @{$\,(\textrm{stat})\,\pm\,$} c @{\,(sys)\,}| c @{$\,\pm\,$} c @{\,(th.sys)\,}| }
\hline
\multicolumn{1}{|c|}{{ \bf H1}} & \multicolumn{3}{c|}{HERA I+II Data} & \multicolumn{2}{c|}{SM} \\
\hline\hline
{\small $\sigma_{\ensuremath{\ell+{P}_{T}^{miss}}}$} &
{\small 0.24} &
{\small 0.05} &
{\small 0.05} &
{\small 0.26} &
{\small 0.04} \\
{\small $\sigma_{W}$} &
{\small 1.23} &
{\small 0.25} &
{\small 0.22} &
{\small 1.31} &
{\small 0.20} \\
\hline
\end{tabular}
\end{center}
\end{table}

\vspace{-0.6cm}

A measurement of the $W$ polarisation fractions is also performed
by H1, as described in \cite{h1wpol}.
Using a 2D fit, optimal values of the left-handed ($F_{-}$) and longitudinal
($F_{0}$) fractions are extracted, as shown in
figure \ref{fig:isolep2} (left) compared to the SM and a FCNC single top model.

\section{Search for Single Top Quark Production}

\begin{figure}
  \includegraphics[height=.47\textwidth]{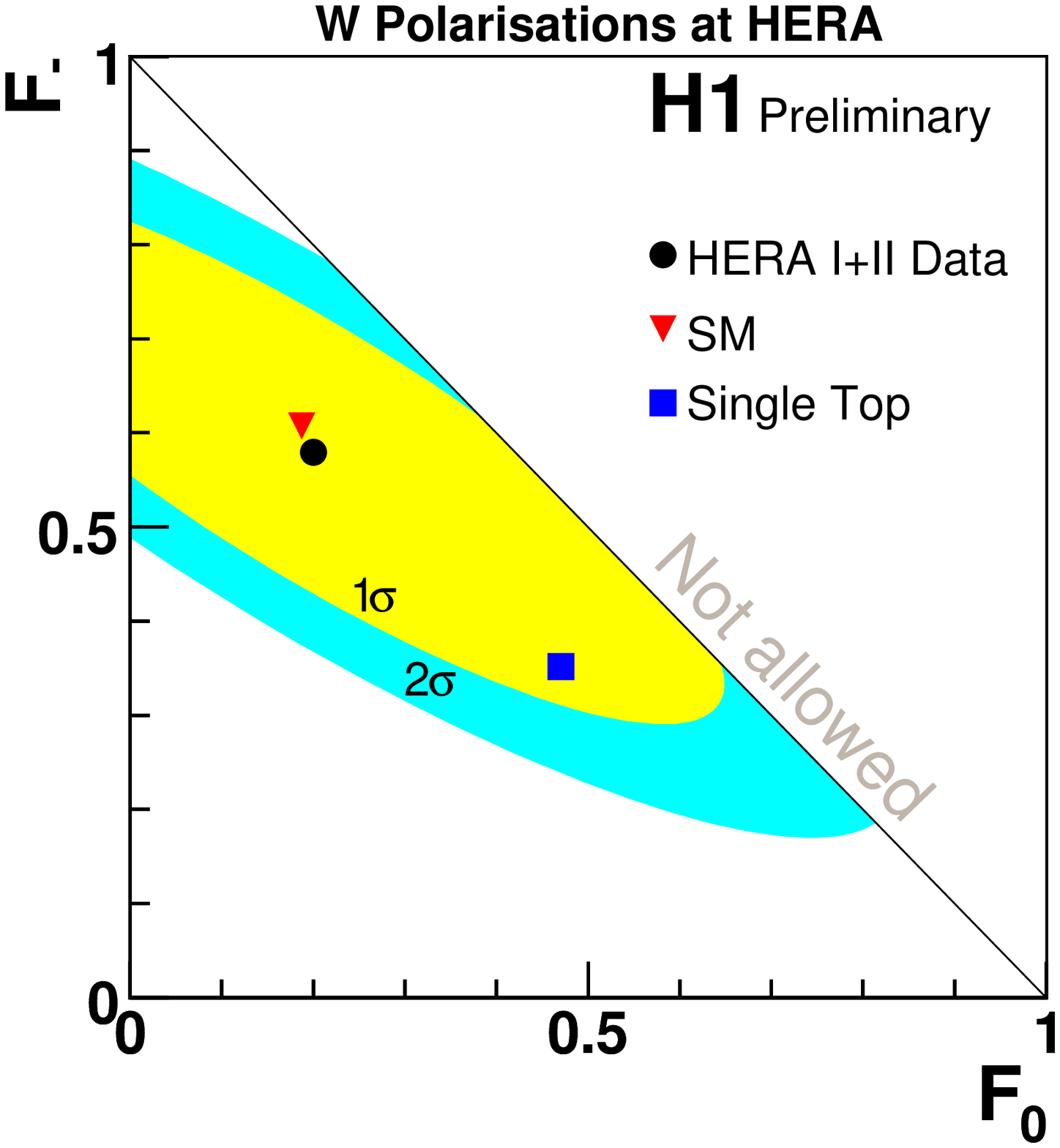}
  \hfill
  \includegraphics[height=.45\textwidth]{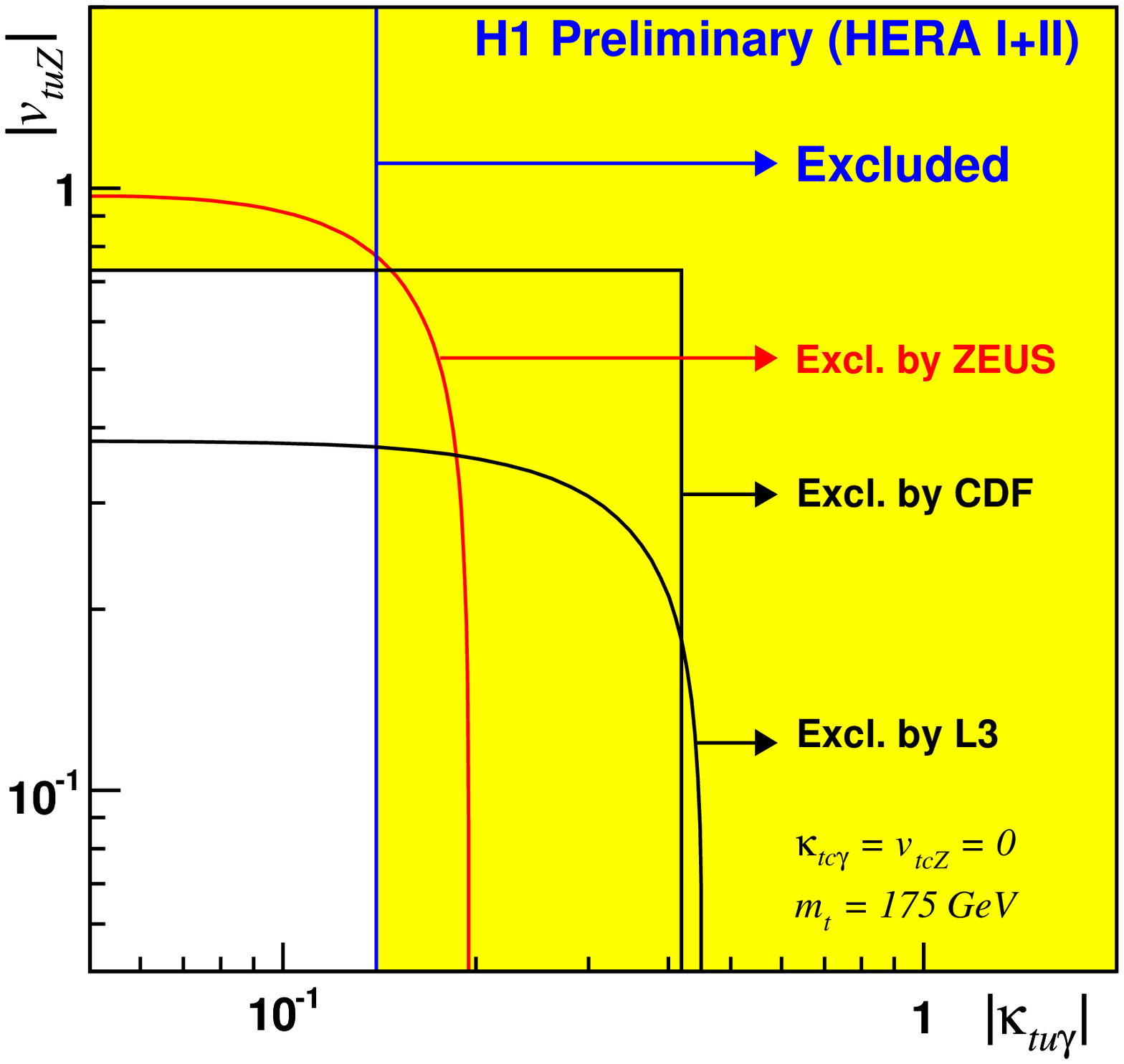}
  \caption{Left: The fit result for the simultaneously extracted left
	handed ($F_{-}$) and longitudinal ($F_{0}$) $W$ boson polarisation
	fractions (point) at 1 and 2$\sigma$ CL (contours). Also shown are
	the values for the SM prediction (triangle) and anomalous single top
	production via FCNC (square). Right: Exclusion limits at the 95\%
	CL in the search for single top production on the anomalous
	$\kappa_{tu\gamma}$ and $v_{tuZ}$ couplings obtained at the
	TeVatron (CDF experiment \cite{cdftop}),
	LEP (L3 experiment \cite{l3top}) and HERA (H1 \cite{h1topnew}
	and ZEUS \cite{zeustop} experiments). Anomalous couplings of the
	charm quark are neglected $\kappa_{tc\gamma}=v_{tcZ}=0$. Limits
	are shown assuming a top mass $m_t=175$~GeV.}
\label{fig:isolep2}
\end{figure}

A search for single top quark production at HERA is performed by H1
as an extension of the search for isolated lepton events,
using the full HERA~I+II $e^{\pm}p$ data \cite{h1top,h1topnew}.
The investigated model considers anomalous production of top quarks in a 
Flavour Changing Neutral Current process involving the coupling
$\kappa_{tu\gamma}$.
%
%
A multivariate analysis is performed to discriminate top from SM
background (dominated by real $W$~production).
No evidence for single top production is observed.
An upper limit on the anomalous top production cross section of
$\sigma_{ep\rightarrow etX}~<~0.16$~pb is established at 95\% CL.
The corresponding H1 limit on the coupling $\kappa_{tu\gamma}~<~0.14$ is
shown in figure \ref{fig:isolep2} (right) and is currently the best limit compared to
those from other colliders\footnote{An improved limit on $v_{tuZ}$ by the CDF Collaboration
was presented at the HEP-EPS 2007 conference; see \cite{cdftopnew}.}.

\begin{table}[t]
\renewcommand{\arraystretch}{1.6}
\begin{center}
  \begin{tabular}{|c|c|c|c|c|}
    \hline
    \multicolumn{2}{|c|}{\large H1+ZEUS Preliminary} &
    Electron &
    Muon &
    Combined \\
    \multicolumn{2}{|c|}{$l$+$P_{T}^{\rm miss}$ events at} &
    obs./exp. &
    obs./exp. &
    obs./exp. \\
    \multicolumn{2}{|c|}{HERA I+II} &
    {\footnotesize (Signal contribution)} &
    {\footnotesize (Signal contribution)} &
    {\footnotesize (Signal contribution)} \\
    \hline
    \hline
    {\footnotesize 1994-2007 $e^{+} p$} &
    {\footnotesize Full Sample} &
    {\footnotesize 39 / 41.3 $\pm$ 5.0 (70\%)}&
    {\footnotesize 18 / 11.8 $\pm$ 1.6 (85\%)}&
    {\footnotesize 57 / 53.1 $\pm$ 6.4 (73\%)}\\
    \cline{2-5}
    {\footnotesize 0.58 fb$^{-1}$} &
    {\footnotesize $P_{T}^{X}~>25$~GeV} &
    {\footnotesize 12 /  7.4 $\pm$ 1.0 (78\%)}&
    {\footnotesize 11 /  7.2 $\pm$ 1.0 (85\%)}& 
    {\footnotesize 23 / 14.6 $\pm$ 1.9 (81\%)}\\
    \hline
    \hline
    {\footnotesize 1998-2006 $e^{-} p$} &
    {\footnotesize Full Sample} &
    {\footnotesize 25 / 31.6 $\pm$ 4.1 (63\%)}&
    {\footnotesize  5 /  8.0 $\pm$ 1.1 (86\%)}&
    {\footnotesize 30 / 39.6 $\pm$ 5.0 (68\%)}\\
    \cline{2-5}
    {\footnotesize 0.39 fb$^{-1}$} &
    {\footnotesize $P_{T}^{X}~>25$~GeV} &
    {\footnotesize  4 /  6.0 $\pm$ 0.8 (67\%)}&
    {\footnotesize  2 /  4.8 $\pm$ 0.7 (87\%)}& 
    {\footnotesize  6 / 10.6 $\pm$ 1.4 (76\%)}\\
    \hline
    \hline
    {\footnotesize 1994-2007 $e^{\pm} p$} &
    {\footnotesize Full Sample} &
    {\footnotesize 64 / 72.9 $\pm$  8.9 (67\%)}&
    {\footnotesize 23 / 19.9 $\pm$  2.6 (85\%)}&
    {\footnotesize 87 / 92.7 $\pm$ 11.2 (71\%)}\\
    \cline{2-5}
    {\footnotesize 0.97 fb$^{-1}$} &
    {\footnotesize $P_{T}^{X}~>25$~GeV} &
    {\footnotesize 16 / 13.3 $\pm$ 1.7 (73\%)}&
    {\footnotesize 13 / 12.0 $\pm$ 1.6 (86\%)}& 
    {\footnotesize 29 / 25.3 $\pm$ 3.2 (79\%)}\\
    \hline
  \end{tabular}
  \caption{Summary of the combined H1+ZEUS results in the search for
	events with isolated electrons or muons and missing transverse
	momentum. Results are shown for the electron and muon channels
	separately as well as combined for the full HERA~I+II $e^{+}p$,
	$e^{-}p$ and $e^{\pm}p$ data. The number of observed data events is
	compared to the SM prediction. The results are shown for the full
	selected sample and for a subsample at large $P_{T}^{X}>25$~GeV.
	The signal component of the SM expectation, dominated by real $W$
	production, is given as a
	percentage in parentheses. The quoted errors contain statistical
	and systematic uncertainties added in quadrature.}
\end{center}
\label{tab:summarytable}
\end{table}

\end{document}